\documentclass[conference]{IEEEtran}
\usepackage{graphicx, epstopdf}
\usepackage{amssymb, amsmath}
\usepackage{commath}
\usepackage{setspace}
\usepackage{acronym}
\usepackage{mathrsfs}
\usepackage{ifthen}
\usepackage{textcomp}
\usepackage{float}
\usepackage{subfigure}
\usepackage{color}
\usepackage{cite}

\IEEEoverridecommandlockouts

\begin{document}
\bstctlcite{ICC09_Ref2:BSTcontrol}

\title{Emergency Route Selection for D2D Cellular Communications During an Urban Terrorist Attack}

\author{Hu Yuan, Weisi Guo, Siyi Wang\textsuperscript{\dag} \\
School of Engineering, University of Warwick, United Kingdom \\
\textsuperscript{\dag}Institute for Telecommunications Research, University of South Australia, Australia \\
hu.yuan@warwick.ac.uk, weisi.guo@warwick.ac.uk, \textsuperscript{\dag}siyi.wang@mymail.unisa.edu.au}

\maketitle

\begin{abstract}
Device-to-Device (D2D) communications is a technology that allows mobile users to relay information to each other, without access to the cellular network.  In this paper, we consider how to dynamically select multi-hop routes for D2D communications in spectrum co-existence with a fully loaded cellular network.  The modelling scenario is that of a real urban environment, when the cellular network is congested during an unexpected event, such as a terrorist attack.  We use D2D to relay data across the urban terrain, in the presence of conventional cellular (CC) communications.

We consider different wireless routing algorithms, namely: shortest-path-routing (SPR), interference-aware-routing (IAR), and broadcast-routing (BR). In general, there is a fundamental trade-off between D2D and CC outage performances, due to their mutual interference relationship.  For different CC outage constraints and D2D end-to-end distances, the paper recommends different D2D routing strategies. The paper also considers the effects of varying user density and urban building material properties on overall D2D relaying feasibility. Over a distance of a kilometre, it was found that the success probability of D2D communications can reach 91\% for a moderate participating user density (400 per square km) and a low wall penetration loss ($<$~10~dB).
\end{abstract}

\section{Introduction}

\subsection{Emergency Communications}

One of the defining trends of our century is the rapid urbanisation in both developed and developing worlds.  Across the planet, more than 50\% of the population now live in cities and this is set to rise rapidly over the next decade~\cite{UN13}.  Modern cities are partly defined by a \emph{high population density} and high mobile phone usage ($>$1 phone per capita).  Therefore, there is an opportunity to achieve multi-hop communications between users. One of the key challenges global cities face is \emph{security from terror attacks}.  Terrorist attacks generally target dense urban areas to deliver the greatest casualty and a high impact.  In the event of such an attack, such as the 9/11 attack in New York City and the 7/7 bombing in London, the wireless communication network becomes overloaded or shutdown.  This is due to the fact that the number of user equipments (UEs) that a base-station (BS) can serve is limited, and the number of radio resource blocks (RRBs) to support services is also limited.  In this paper, we assume the cellular network is fully loaded with traffic, and a large set of UEs are seeking alternative ways to relay vital data. Device-to-Device (D2D) communications is a way of allowing UEs to act as relays for each other.  The BSs of the cellular network are avoided in terms of data-bearing channels, but may or may not serve as a coordinator or facilitator to D2D channels.  In this paper, we treat D2D channels as emergency data channels, whereby the \emph{end-to-end outage performance} is of greatest importance.

\subsection{Interference Aware Routing Algorithms}

Routing in wireless multi-hop communications is a well addressed research area.  Dynamic spectrum analysis routing has been proposed and analysed in~\cite{Goldsmith10}, so that relaying and other conventional communication links can co-exist.  In \emph{orthogonal frequency relaying}, existing multi-hop schemes focus on an intuitive shortest-path-routing (SPR) analysis, and attempts to maximize the performance through cooperative transmission and interference cancellation techniques. In particular, cooperative multi-hop communications on orthogonal channels has been well investigated.  For example, research in~\cite{Chang13} has shown that a collaborative cluster of D2D UEs can also achieve significant energy saving, or alternatively a transmission range extension with the same transmit energy budget.  Similarly, our own work in~\cite{Guo11JSAC, Guo10VTCS} found that under a fixed energy budget, increased cooperation does not monotonically lead to increased transmission reliability.  The relationship is in fact convex, and for any given system setup and channel conditions, there exists an optimal set of cooperation partners which maximises the transmission reliability.  Other research schemes use coordinated transmission and MIMO technologies to control and cancel interference between D2D channels and cellular channels~\cite{Chien12, Min11}.

For a multi-hop network that mutually interferes with another co-frequency overlay network (i.e., an umbrella cellular network), the problem of \emph{interference aware route selection} is not well researched.  One example of interference aware route selection has recently been studied in~\cite{Parissidis11}, where an artificial interference concept was introduced in terms of circular zones. Clearly, this concept has limitations in the context of realistic urban environments, where the transmission range of a signal is not uniform.  This paper introduces a novel interference-aware routing algorithm for emergency transmission in urban environments.  Mathematical models using stochastic geometry are presented for conventional cellular (CC) and D2D communications.  Several routing strategies' algorithms and performances are compared and results for an urban environment with varying UE densities and building materials are presented.
\begin{figure}[t]
    \centering
    \includegraphics[width=1.00\linewidth]{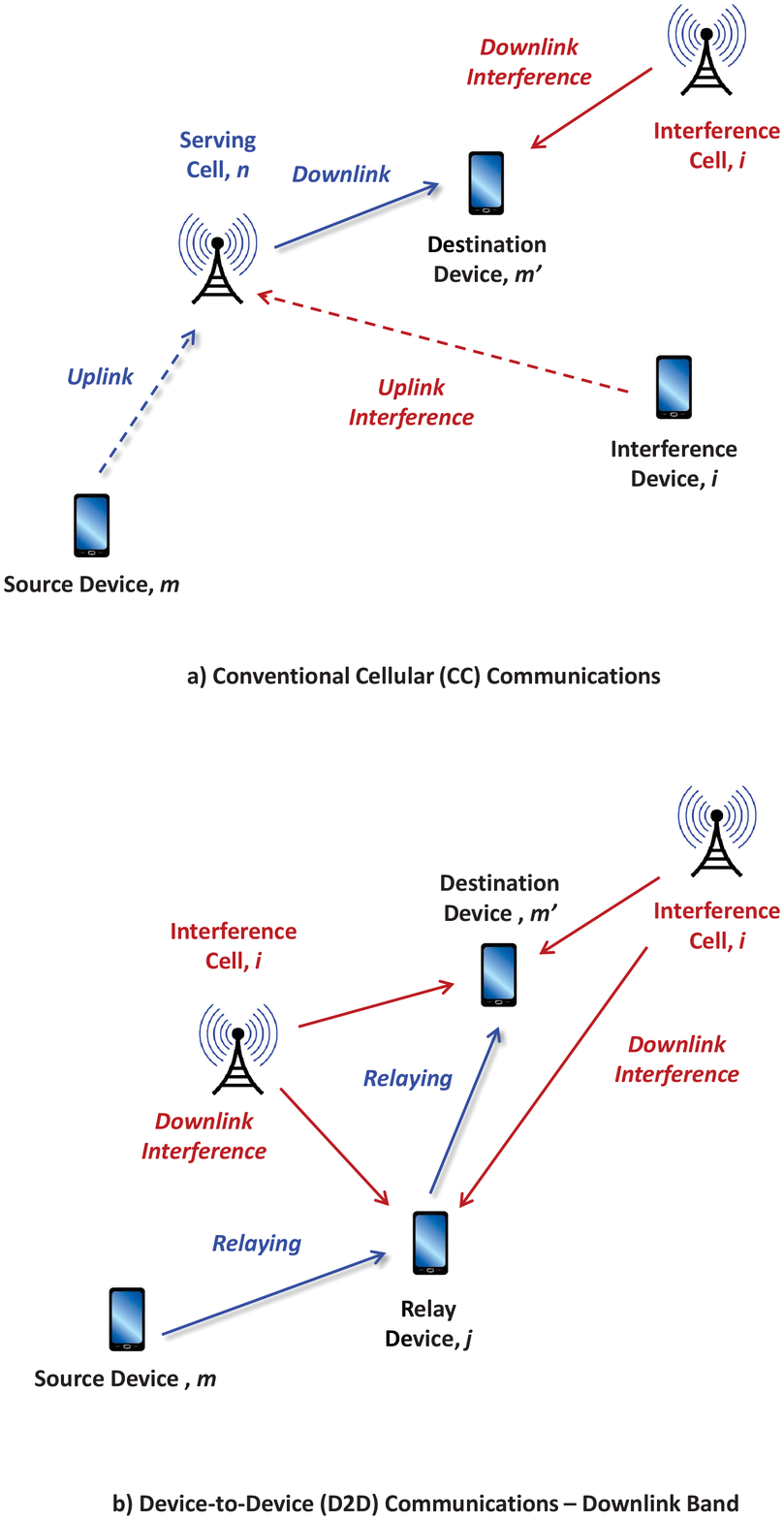}
    \caption{Cellular System Setup: (a) Conventional Communications (CC) between two UEs with interference from neighbouring BSs; (b) Device-to-Device (D2D) emergency multi-hop communications with interference from neighbouring BSs.}
    \label{System}
\end{figure}

\section{Experiment Setup}

\subsection{4G Cellular Network}

The system considered is illustrated in Fig.~\ref{System}, which is an OFDMA based multiple-access network such as 4G LTE. It consists of a number of static macro BSs and UEs.  In this paper, we consider the communications between 2 arbitrary UEs, routing data via one of two ways: Conventional Cellular (CC) channels, or Device-to-Device (D2D) channels.  We illustrate our idea with UEs in the same BS's coverage area, but the idea is easily extendible to UEs across multiple BSs' coverage areas.

We explain the 2 different transmission modes in greater detail and note that they operate in co-existence:
\begin{itemize}
  \item CC: the source UE transmits data to the serving-BS using the uplink band and the destination UE receives data from the same or different serving-BS in the downlink band, as shown as Fig.~\ref{System}a.
  \item D2D: the source UE transmits data to the relaying UEs and the destination UEs using a band (downlink or uplink), and the interference at each UE is from neighbouring BSs (this may or may not include the parent BS), as shown in Fig.~\ref{System}b.
\end{itemize}

In terms of the physical layer, the system utilises real modulation-and-coding schemes (MCS) for 4G LTE, which comprises of 27 MCS combinations~\cite{ViennaPhy10}.  The minimum signal-to-interference noise ratio (SINR) required for data flow is -6~dB in the urban environment.  A full list of experimental parameters and corresponding values is found in Table~\ref{Parameters}. In terms of the \emph{traffic load}, every BS in the cellular network experiences a full buffer traffic from CC sources during the aftermath of a terrorist attack.  Furthermore, UEs that wish to communicate to each other need to share the same spectrum and use D2D multiple relaying.  Therefore, the dominant issue is the mutual interference from CC and D2D channels in co-existence.
\begin{figure}[t]
    \centering
    \includegraphics[width=1.00\linewidth]{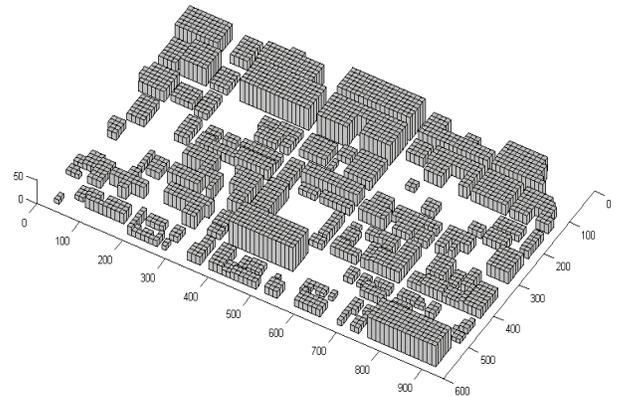}
    \caption{3D building model of a section in Ottawa city created for propagation modelling.}
    \label{Propagation}
\end{figure}

\subsection{Urban Propagation Environment}

The propagation environment used in this study is a real city centre in Ottawa City in Canada.  A 0.92~km $\times$ 0.55~km grid is selected that comprises of approximately 80 buildings of various shapes and dimensions.  The streets are generally orthogonal and follow a classical Manhattan model layout~\cite{3GPP10Channel}.  Specifically the propagation model used is the Urban Micro (UMi) model in 3GPP.  The Line-of-Sight (LOS) and Non-LOS (NLOS) is determined by ray-tracing in the 3D city model shown in Fig.~\ref{Propagation}.  We assume all UEs are \emph{outdoors} in the event of a terrorist attack, but communication signals can go through buildings.  The penetration loss as a result of indoor-to-outdoor and outdoor-to-indoor propagation is adjustable as a function of building material properties.
\begin{table}[t]
    \caption{Experimental and Theoretical Parameters}
    \begin{center}
        \begin{tabular}{|l|l|}
          \hline
          \emph{Parameter}                  & \emph{Value}                      \\
          \hline
          Bandwidth                         & 20 MHz                            \\
          Transmit Frequency                & 2.1 GHz                           \\
          Propagation Model                 & 3GPP UMi                          \\
          Simulation Area                   & 0.51 km\textsuperscript{2}                     \\
          UE Distribution                   & Random Outdoors                   \\
          BS Density                        & $\Lambda_{\mathrm{BS}}$           \\
          Minimum SNR for Data              & $\zeta=-6$~dB                     \\
          AWGN Power                        & $-162$~dB                           \\
          BS Antenna Height                 & 45 m                              \\
          BS Transmit Power                 & 40 W                              \\
          D2D UE Transmit Power             & 0.1 W                             \\
          D2D Source-Destination Dist.      & 0.45--0.9 Cell Diameter         \\
          D2D UE Density                    & 0--400 per sq. km                 \\
          Wall Penetration Loss             & 5--30~dB                          \\
          Traffic Model                     & Full Buffer                       \\
          Multi-path Fading                 & Rayleigh                          \\
          Shadow Fading Variance            & 6~dB                              \\
          \hline
        \end{tabular}
    \end{center}
    \label{Parameters}
\end{table}

\section{Outage Probability Analysis}

The paper utilises a combination of theoretical framework and Monte-Carlo simulation results to validate our investigation.  This section now introduces the theoretical framework, which also sheds light on the underlying mechanics of the 2-tier system.  The paper considers 2 arbitrary UEs, which have an end-to-end distance of $r_{m,n}$ and $r_{n,m'}$ to their serving BS respectively.  The instantaneous SINR of a communication link from $n$ to $m$ is defined as:
\begin{equation}\begin{split} \label{SINR}
\gamma_{n,m} = \frac{h_{n,m}P_{n,m}\lambda r_{n,m}^{-\alpha}} {\sigma^{2} + \sum\limits_{\substack{i\in\Phi\\i\neq n}} h_{i,m}P_{i,m}\lambda r_{i,m}^{-\alpha}},
\end{split}\end{equation} where $W$ is the AWGN power, $h$ is the fading gain, $P$ is the transmit power, $\lambda$ is the frequency dependent pathloss, and $r$ is the distance.  There is a set of $\Phi$ interferers, and it can be assumed that for an interference-limited network, the AWGN power is negligible.

For CC communications, the end-to-end outage probability (SINR falling below $\zeta$) of UE $m$ communicating to UE $m'$ is given as a function of the uplink and downlink outage probabilities:
\begin{equation}\begin{split} \label{OutageCC0}
\mathrm{P}_{\mathrm{CC,out}}(m,m') = 1 - \mathbb{P}(\overline{\gamma}_{m,n} > \zeta) \mathbb{P}(\overline{\gamma}_{n,m'} > \zeta).
\end{split}\end{equation}

For downlink transmission, the interference arrives from adjacent BSs with a spatial density of $\Lambda_{\mathrm{BS}}$.  For uplink transmission, the interference arrives from other UEs in adjacent BSs.  Elementary stochastic geometry can be utilised to yield the probability of successful transmission in the downlink channels~\cite{Wang12HPCC, Haenggi13}:
\begin{equation}\begin{split} \label{OutageCC}
\mathrm{P}_{\mathrm{CC,out}}(m,m') = 1 - \exp\bigg[-\Lambda_{\mathrm{BS}} \pi \left(r_{m,n}^{2}+r_{n,m'}^{2}\bigg) \mathcal{A}(\zeta,4)\right],
\end{split}\end{equation} where the $\mathcal{A}()$ function is given by:
\begin{equation}\begin{split} \label{QFunction}
\mathcal{A}(\zeta,\alpha) &= \int_{\zeta^{-2/\alpha}}^{+\infty} \frac{\zeta^{2/\alpha}}{1+u^{\alpha/2}}  \, \dif u, \\
&= \sqrt{\zeta}\arctan({\sqrt{\zeta}}) \quad \mbox{for: } \alpha = 4.
\end{split}\end{equation}
The uplink channel analysis is beyond the scope of this paper.
\begin{figure*}[t]
    \centering
    \includegraphics[width=0.95\linewidth]{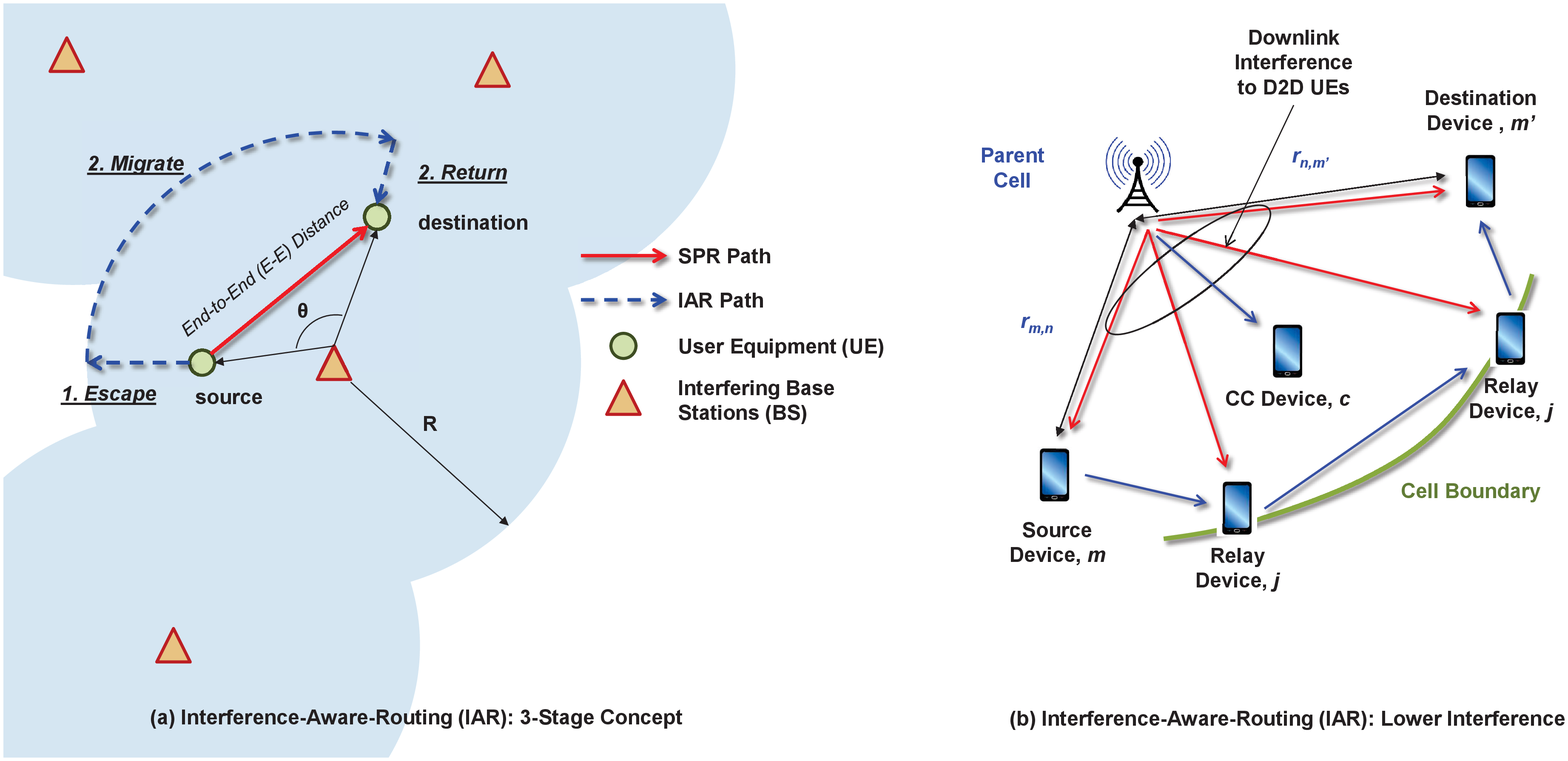}
    \caption{Interference-Aware-Routing (IAR): (a) 3-Stage Process; (b) Reduced Interference.}
    \label{MinImpact}
\end{figure*}

For D2D communications, the paper considers additional UEs that cannot be scheduled radio resources to transmit their data.  In any transmission band, the outage probability for non-cooperative decode-and-forward (DF) relaying is given as a function of the product of the success probability for each link:
\begin{equation}\begin{split} \label{OutageD2DUL}
\mathrm{P}_{\mathrm{D2D,out}} &= 1 - \prod_{j=1}^{J} \bigg( 1- \mathbb{E}_{R}\left(\mathrm{P}_{\mathrm{D2D-SPR,out}}\right) \bigg) \\
\end{split}\end{equation} where the total number of hops $J$ is determined by the density of UEs in the network, the distance between the source and destination UEs, and the route selected.  Further expansion of the expression is beyond the scope of the paper.

\section{D2D Routing Strategies}

\subsection{Shortest-Path-Routing (SPR) and Broadcast-Routing (BR) Algorithm}

In Shortest-Path-Routing (SPR), each D2D UE knows its location through GPS and other wireless localisation means (i.e., wireless fingerprinting and triangulation).  The paper now outlines the step-by-step D2D algorithm needed to achieve shortest path routing from a generic UE pair ($m$ to $m'$).  Assuming that SPR is chosen as the routing strategy, the multi-hop algorithm works in the following manner:
\begin{enumerate}
  \item Source UE $m$ is able to detect which of its neighbouring UEs it can successfully transmit to with some arbitrary outage probability threshold $\zeta$ that it needs to satisfy;
  \item Given a selection of potential relay UEs $j$, it is able to select one closest to the final destination UE $m'$;
  \item This process is repeated until the destination UE $m'$ is reached.
\end{enumerate}
Whilst D2D transmissions are taking place, the regular CC channels will suffer additional interference.  The network effectively becomes a 2-tier co-band network in the DL band and the outage probability of CC in \eqref{OutageCC} needs to be revised.

In Broadcast-Routing (BR), each D2D UE broadcasts its data, which may or may not be received by a number of other UEs.  Other UEs simply continue to broadcast this data.  Therefore, a propagative ripple effect in the data exists across the network.  It is likely that the interference caused to other CC channels will be the greatest in this scheme.

\subsection{Interference Aware Routing (IAR) Algorithm}

The idea behind IAR is to reduce the D2D interference caused to the BS received in the uplink band.  This is intuitively achieved if the D2D routing process occurs along the BS's cell boundary, where the distance to adjacent BSs is maximised and the aggregate interference to adjacent BSs is minimised.  The IAR path has 3 distinct stages (Fig.~\ref{MinImpact}a):
\begin{itemize}
  \item \emph{Stage 1 (Escape to Cell Boundary):} from source UE $m$ to closest boundary UE $j$;
  \item \emph{Stage 2 (Migrate along Cell Boundary):} from boundary UE closest to the source to a boundary UE closest to the destination;
  \item \emph{Stage 3 (Return from Cell Boundary):} from the boundary UE closest to the destination to the destination UE $m'$.
\end{itemize}
Each stage of the IAR actually utilises the SPR algorithm.  Clearly the route is longer than the SPR path, but the advantages are that the interference from CC UEs can be reduced significantly due to the increased distance from the parent-BS.  This is illustrated in Fig.~\ref{MinImpact}b for the downlink (DL) channel.  A similar case is true for the uplink (UL) channel, which is beyond the scope of this paper.  Whilst the D2D route is closer to other interfering BSs, the combined interference effect across all BSs is reduced.  The corresponding uplink interference scenario is not illustrated in this paper, but it is considered in the results section.
\begin{figure*}[t]
    \centering
    \includegraphics[width=0.80\linewidth]{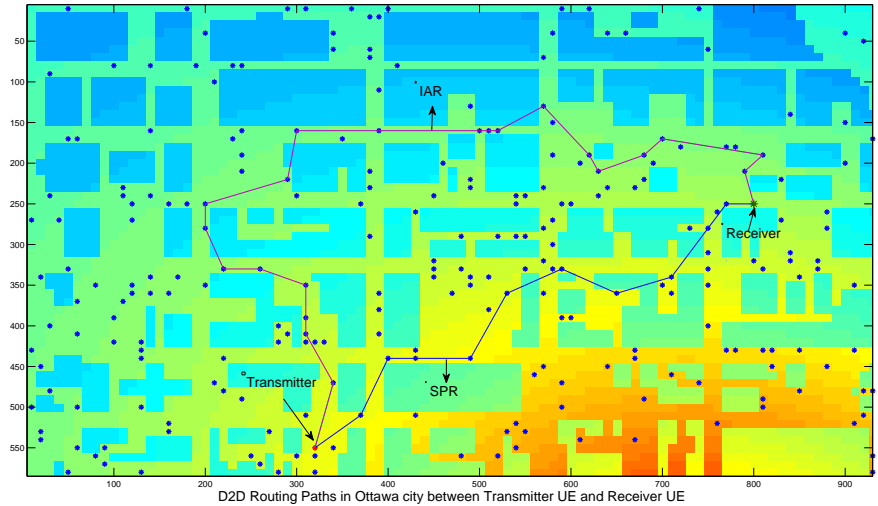}
    \caption{Simulated D2D Routing Paths in Ottawa city between transmitter UE (Tx) and Receiver UE (Rx) for Shortest-Path-Routing (SPR) and Interference-Aware-Routing (IAR).  The diagram is under-laid with the interference power received at each location.  Stars represent outdoor UE positions.}
    \label{Propagation3}
\end{figure*}

\section{Results and Analysis}

\subsection{D2D Performance}

The paper first examines the feasibility of D2D routing when the base station is fully loaded.  In Fig.~\ref{Propagation3}, the simulation results show the simulated end-to-end D2D routing paths between an arbitrary transmitter UE (Tx) and receiver UE (Rx) for both Shortest-Path-Routing (SPR) and Interference-Aware-Routing (IAR) in Ottawa city.  The first observation is that the IAR path is approximately 35\% longer than the SPR path in this particular case, and this value reflects the average as well.  However, the IAR path mostly travels in the low interference power regions (green to light blue), whereas the SPR path travels in the high interference power regions (yellow).  Therefore, the mutual interference between the IAR D2D UEs and the CC UEs is lower than the SPR case.  Figure~\ref{Propagation3} also shows the downlink SINR of CC links in the Ottawa city centre.
\begin{figure}[t]
    \centering
    \includegraphics[width=0.95\linewidth]{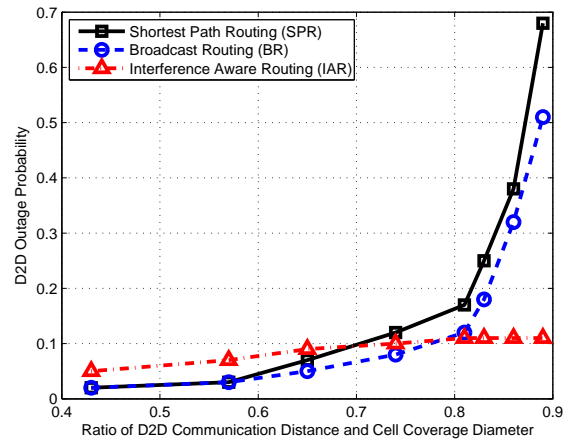}
    \caption{D2D outage probability as a function of the ratio between D2D distance and cell coverage diameter.}
    \label{OutageRouting}
\end{figure}
\begin{figure}[t]
    \centering
    \includegraphics[width=1.00\linewidth]{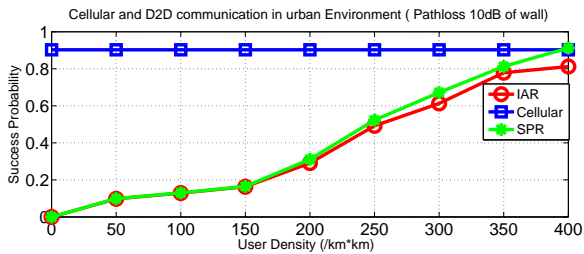}
    \caption{D2D routing success probability as a function of D2D UE density.}
    \label{Results}
\end{figure}
\begin{figure}[t]
    \centering
    \includegraphics[width=1.00\linewidth]{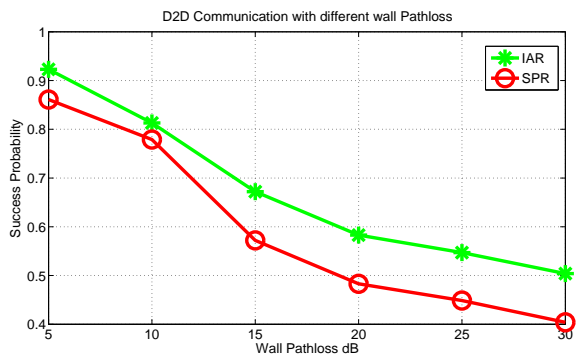}
    \caption{D2D routing success probability as a function of building outer wall penetration loss (dB).}
    \label{Results2}
\end{figure}

\subsubsection{D2D Routing Distance}

Fig.~\ref{OutageRouting} compares the routing algorithms: i) SPR, ii) BR, and iii) IAR, all using downlink (DL) bands.  It is found that for small D2D communication distances, both SPR and BR achieve lower outage probabilities than IAR.  This is intuitive as the IAR routing algorithm stipulates that even when communicating short distances, the route must escape to the cell edge and return.  The increase in route distance is likely to be several folds higher than the SPR and BR cases.

Whilst BR achieves a slightly better performance than SPR, the interference it causes to CC UEs is more significant as more transmissions are required.  For D2D communication distances that are significantly greater than the cell radius, there is a high probability that the BR and SPR paths will pass near the BS.  This will cause significant interference between CC links (via the BS) and D2D links.  The IAR mechanism allows the routing to avoid the BS' site location and maximise the mutual distance between the D2D multi-hop path and the BS.  This reduction in mutual interference leads to an improved overall performance, despite increasing the overall hop length.

\subsubsection{User Density}

Fig.~\ref{Results} shows the D2D routing success probability as a function of D2D UE density, varying from 0 to 400 UEs per square km.  The success probability rises to over 80\% when the UE density is over 400/km\textsuperscript{2} and the results for IAR and SPR are remarkably similar. That is to say, IAR is just as effective as SPR, whilst minimizing interference to CC UEs in the centre regions of the cell's coverage area.

\subsubsection{Wall Penetration Loss}

Fig.~\ref{Results2} shows the D2D routing success probability as a function of the building outer wall penetration loss (dB), varying from 5 to 30~dB.  The success probability falls to below 50\% when the building outer wall penetration loss is at 30~dB (thick wall).  The IAR performs consistently better than SPR for this set of results by up to 10\%.  Therefore, D2D is possible under certain environmental and user density scenarios.  More specifically, when the co-network UE density is over 400/km\textsuperscript{2} and when the walls in the city are not very thick (less than 10~dB).
despite increasing the overall hop length.
\begin{figure}[t]
    \centering
    \includegraphics[width=0.95\linewidth]{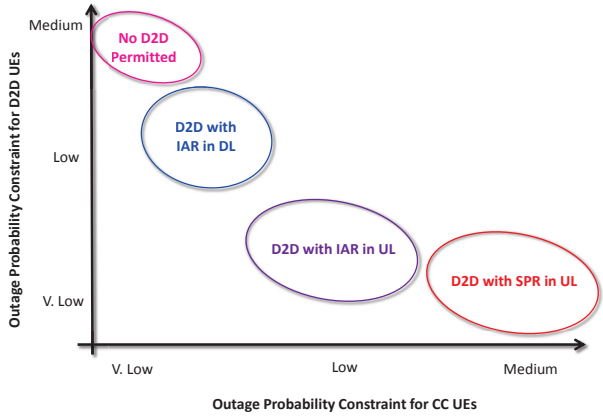}
    \caption{Dynamic D2D band- and routing-strategy, based on CC outage probability constraint.}
    \label{Final}
\end{figure}

\subsection{Under CC Performance Constraint}

One of the key advantages of IAR routing over SPR routing is that it reduces the interference emitted to regular CC UEs.  By picking a routing path that travels predominantly along the traditional cell-edge, it maximizes the distance to the majority of CC UEs.  The paper now expands the IAR routing to both consider uplink (UL) and downlink (DL) bands.

Fig.~\ref{Final} shows the D2D outage probability for various CC outage constraints.  The results show that there is an intuitive trade-off in outage probability between CC and D2D UEs.  For a stringent CC outage constraint, D2D transmission is not permitted.  As the CC constraint gets relaxed, the D2D routing method changes from IAR to SPR, and from the DL to the UL band.  More specifically, the results show that for:
\begin{itemize}
  \item CC outage constraint $<5\%$: no D2D is permitted;
  \item CC outage constraint $<12\%$: D2D using IAR in DL can achieve the lowest outage probability of 20\%;
  \item CC outage constraint $<15\%$: D2D using IAR in UL can achieve the lowest outage probability of 8\%;
  \item CC outage constraint $<40\%$: D2D using SPR in UL can achieve the lowest outage probability of 3\%;
\end{itemize}

There is an intuitive trade-off in outage probability between CC and D2D, what has been improved is that by dynamically selecting the D2D routing method and transmission band, the D2D outage can be minimised.  The D2D transmit band that causes the least interference to CC is the DL band, but the D2D outage is reasonably high.  As the outage constraint is relaxed in CC, there is a shift from interference aware transmit band and routing paths, to the shortest path in UL band.

\section{Conclusions}

Device-to-Device (D2D) communications is a technology that allows mobile user equipments to relay information to each other, without data access to the cellular network.  In this paper, we assume there has been a terrorist attack to a real city and that the cellular network is congested.  Emergency D2D communications needs to co-exist with the conventional cellular (CC) communications.  The paper shows that in such a co-existence and mutually interfering scenario, interference-aware-routing (IAR) is superior to the intuitive shortest-path-routing (SPR) and broadcasting algorithms, if the overall transmission range is over 80\% of a cell's coverage diameter.  Otherwise, for short distance D2D communications, the SPR and BR algorithms perform better.  In general, there is a fundamental trade-off between D2D and CC outage performances, due to their mutual interference.  For different CC outage constraints and D2D distances, the paper shows how different D2D routing strategies should be selected.  In terms of D2D feasibility, the results show that the D2D emergency channel can achieve up to a high success communication probability of 91\% when the user density is high (400 available users per square km), but can drop to 50\% when the user density falls or when the building's wall penetration loss is relatively high (30~dB). Therefore, there remains significant challenges related to whether D2D communications in urban areas is feasible in the event of an emergency that overloads the cellular network.  \\

\section*{Acknowledgement}
The work in this paper is partly supported by the British Council Knowledge Transfer Partnership, and the University of Warwick's Engineering Scholarship scheme. \\

\bibliographystyle{IEEEtran}
\bibliography{IEEEabrv,Bibliography}

\end{document}